\documentclass[aps,pra,twocolumn,superscriptaddress]{revtex4-2}
\usepackage{bm}
\usepackage{amsmath}
\usepackage{amsfonts}
\usepackage{amssymb}
\usepackage{graphicx}
\usepackage{color}  
\usepackage{xcolor}
\usepackage[braket, qm]{qcircuit}   
\usepackage[colorlinks,linkcolor=blue,anchorcolor=blue,citecolor=blue,urlcolor=blue]{hyperref}
\usepackage{dcolumn}

\begin{document}


\title{Variational quantum compiling for three-qubit gates design in quantum dots}

\author{Yuanyang Zhou}
\affiliation{Institute for Quantum Science and Technology, Department of Physics, Shanghai University, Shanghai 200444, China}

\author{Huaxin He}
\affiliation{Institute for Quantum Science and Technology, Department of Physics, Shanghai University, Shanghai 200444, China}

\author{Fengtao Pang}
\affiliation{Institute for Quantum Science and Technology, Department of Physics, Shanghai University, Shanghai 200444, China}

\author{Hao Lyu}
\affiliation{Quantum Systems Unit, Okinawa Institute of Science and Technology Graduate University, Onna, Okinawa 904-0495, Japan}

\author{Yongping Zhang}
\email{yongping11@t.shu.edu.cn}
\affiliation{Institute for Quantum Science and Technology, Department of Physics, Shanghai University, Shanghai 200444, China}

\author{Xi Chen}
\email{xi.chen@csic.es}
\affiliation{Instituto de Ciencia de Materiales de Madrid (CSIC), Cantoblanco, E-28049 Madrid, Spain}

\date{\today}

\begin{abstract}


Semiconductor quantum dots offer a promising platform for controlling spin qubits and realizing quantum logic gates, essential for scalable quantum computing. In this work, we utilize a variational quantum compiling algorithm to design efficient three-qubit gates using a time-independent Hamiltonian composed of only physical interaction terms. The resulting gates, including the Toffoli and Fredkin gates, demonstrate high fidelity and robustness against both coherent and incoherent noise sources, including charge and nuclear spin noise. This method is applicable to a wide range of physical systems, such as superconducting qubits and trapped ions, paving the way for more resilient and universal quantum computing architectures.

\end{abstract}

\maketitle

\section{Introduction}

Quantum computing has garnered significant attention in recent years due to its potential to address complex problems that are intractable for classical computers~\cite{Preskill,farhi2001quantum,ladd2010quantum,RevModPhys.94.015004,cerezo2021variational}. Compared to classical computers, quantum computers promise exponential speedups for problems such as large-number factorization~\cite{shor1994algorithms}, quantum many-body simulations~\cite{lloyd1996universal}, and solving certain linear systems~\cite{harrow2009quantum}. Additionally, they offer quadratic speedups for tasks like unstructured search~\cite{grover1996fast}. 
The performance of quantum computers and algorithms fundamentally depends on the precision of quantum gates, which manipulate qubits and serve as the building blocks of quantum circuits. These high-precision gates are essential for realizing reliable and scalable quantum computation.

With rapid advances in quantum technology, quantum computing protocols have been implemented across various platforms, including Rydberg atoms, trapped ions and superconducting qubits, and spin qubits in semiconductor quantum dots~\cite{nielsen2001quantum}. Among these, quantum dots~\cite{kloeffel2013prospects,zhang2019semiconductor,zajac2018resonantly,huang2019fidelity,watson2018programmable,hanson2007spins} have emerged as a particularly promising platform for quantum computing, owing to their scalability, flexibility, and the well-established maturity of lithography technology~\cite{dodson2020fabrication,hansen2022implementation}. Recently, high-fidelity single-qubit and two-qubit gates have been experimentally demonstrated in quantum dots~\cite{yoneda2018quantum,yang2019silicon,noiri2022fast,xue2022quantum,weinstein2023universal}. However, the three-qubit Toffoli gate, which is essential for reversible classical computing and quantum error correction, remains challenging to implement with high fidelity in quantum dots due to charge noise~\cite{van2019impact,huang2018spin}. Traditional numerical optimization methods, such as GRAPE and CRAB, struggle to optimize control fields in the presence of exchange interactions and single-qubit driving~\cite{khaneja2005optimal,caneva2011chopped}.
Machine learning algorithms have been proposed to design robust two-qubit and three-qubit gates in noisy environments~\cite{PhysRevApplied.6.054005,An_2019,kanaar2022two,gungordu2022robust,kanaar2024neural,PhysRevA.110.032614}. However, these approaches often require controlling different Hamiltonians over specific time periods. The associated dynamical control can introduce additional noise, reducing gate fidelity and making such methods impractical for current quantum hardware~\cite{rimbach2023simple}.
A promising alternative is to minimize noise from control pulses by employing time-independent Hamiltonians. This approach can significantly reduce gate execution time while enhancing computational accuracy and scalability, making it a practical solution for implementing large-scale quantum gates.

To optimize time-independent Hamiltonians, it is typically necessary to iteratively solve a set of nonlinear equations to determine the appropriate parameters~\cite{banchi2016quantum,innocenti2020supervised,lewis2024geodesic}. While gradient-based optimization methods are commonly used \cite{PhysRevResearch.6.013147}, they are often prone to getting stuck in local minima or saddle points, especially when the objective function is highly non-linear or contains multiple local optima. Variational quantum algorithms (VQAs)~\cite{farhi2014quantum,RevModPhys.92.015003}, emerging as suitable algorithms for NISQ quantum devices, offer a way to overcome this limitation, as demonstrated in prior works
\cite{PhysRevX.7.021027,PRXQuantum.2.010101}. Motivated by these, we apply the quantum-assisted quantum compilation (QAQC) algorithm~\cite{khatri2019quantum} to generate three-qubit quantum gates in a time-independent quantum-dot system. First, we consider a noiseless environment to determine the optimal control pulses. This approach significantly reduces the computational cost associated with evaluating the loss function, enabling more efficient optimization. As a result, we achieve a gate fidelity exceeding 99.99\%. Next, we optimize the control pulses in the presence of coherent noises (e.g., charge noise and nuclear noise), ensuring robustness within certain thresholds. Finally, we demonstrate that high-fidelity quantum gates can also be constructed in the presence of incoherent noises (e.g., quantum device noise).

The remainder of the paper is outlined as follows. In Sec.~\ref{model}, we introduce the model, Hamiltonian and method. In Sec.~\ref{result} and \ref{discussion}, we present the preparation of the Toffoli and Fredkin gates 
and discuss the influence of environments, with coherent and incoherent noise. Finally, the conclusion is provided in Sec.~\ref{conclusion}.

\section{Model, Hamiltonian, and method}
\label{model}

\begin{figure*}[htbp]
\includegraphics[width=0.98\textwidth]{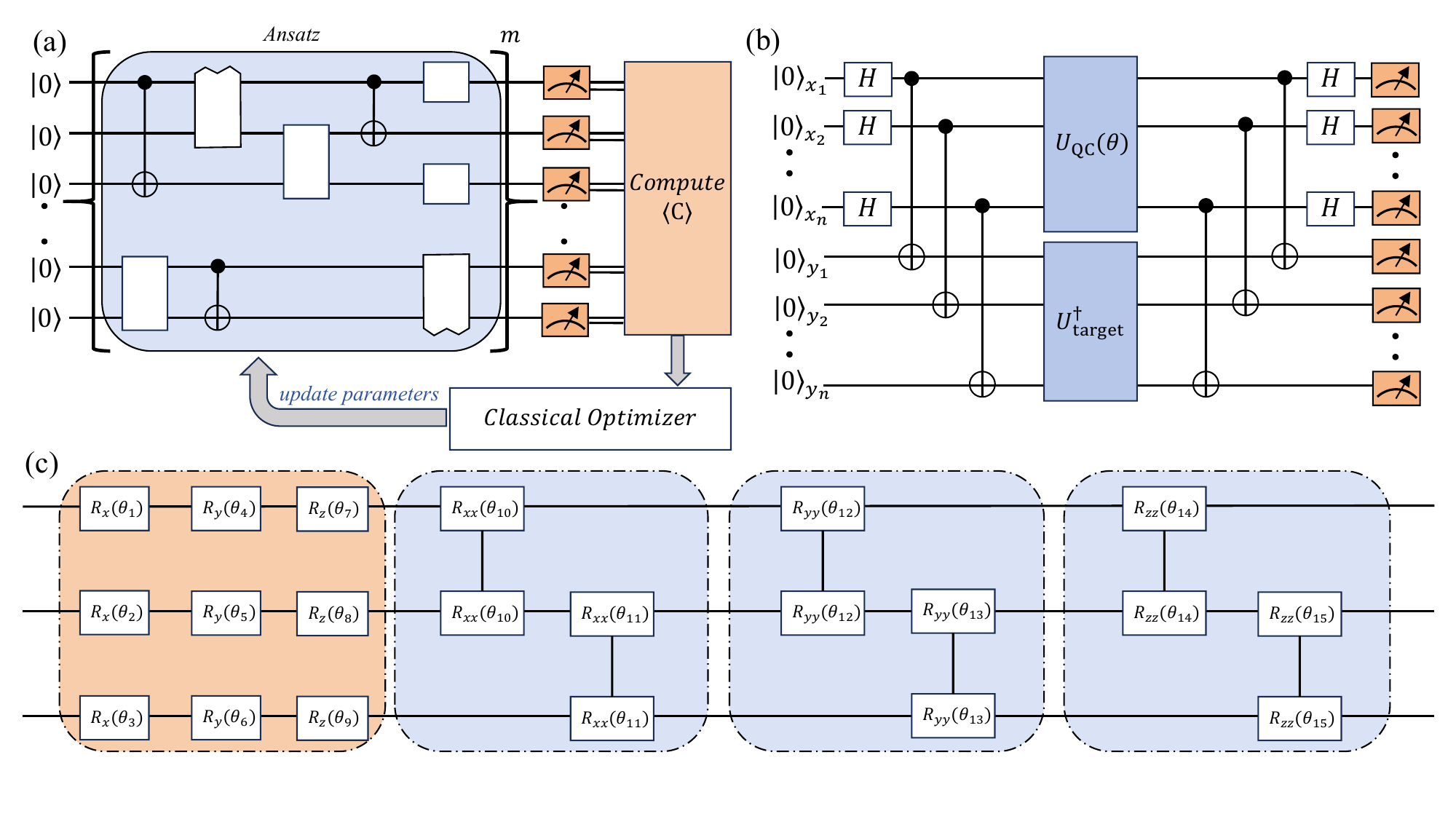}
\caption{
(a) Sketch of quantum circuits for VQA. The structure of the VQA begins with the preparation of an initial state, followed by the application of a parameterized ansatz circuit. The evolved state from this circuit is used to evaluate the cost function, which quantifies the difference between the current state and the target solution. This cost function is minimized iteratively using a classical optimizer, which updates the parameters of the ansatz circuit.
(b) The Hilbert-Schmidt (HS) test. To evaluate the cost function efficiently on a quantum processor, the Hilbert-Schmidt test is employed. Two unitary operators $U_{\text{QC}}(\bm{\theta})$ and $U^\dagger_{\text{target}}$ act sequentially on $n$-qubit basis states. The resultant operator  $U^\dagger_{\text{target}}U_{\text{QC}}(\bm{\theta})$ is applied to the system, and the probability of  measuring all $2n$-qubits in the $\ket{0}$ state is recorded. This probability directly correlates with the cost function [Eq.~\eqref{cost}], enabling its efficient computation on quantum hardware.
(c) Hamiltonian variational ansatz circuit. For a system of size $N=3$ and a single Trotter-Suzuki layer ($m=1$), the HVA circuit is depicted. The circuit comprises single-qubit rotation gates (orange) and entangling gates (blue) between pairs of qubits. This layered structure facilitates the parameterized simulation of the target Hamiltonian's time evolution, with each gate contributing to the flexible encoding of the solution space.}
\label{fig1}
\end{figure*}

We consider a chain of three exchange-coupled spin qubits confined in quantum dots, 
where intrinsic spin-orbit interactions induce anisotropic spin-spin interactions~\cite{geyer2024anisotropic}. 
External magnetic fields can be applied to tune the energies and spin rotations of the spin qubits.
The system is described by the anisotropic Heisenberg model ($\hbar \equiv 1$)~\cite{PhysRevA.81.052331,PhysRevLett.93.140501,PhysRevLett.104.126401},
\begin{equation}
H = \sum^3_{i=1}\sum_{\alpha=x,y,z} h_i^{\alpha} \sigma^{(i)}_{\alpha}+ \sum_{i=1,2} \sum_{\alpha=x,y,z} J_{i}^{\alpha} \sigma^{(i)}_\alpha \sigma^{(i+1)}_\alpha,
\label{Hamiltonian}
\end{equation}
where $\sigma_{\alpha}^{(i)}$ ($i=1,2,3$) are the standard Pauli operators for the spin in the $i$-th quantum dot. The spin-splitting energies are given by
$h_i^{\alpha} = \mu_B g_i B^\alpha_i$ ($\alpha=x,y,z$), which are tunable via tuning the local magnetic field $\bm{B}_i = (B^x_{i} ,B^y_{i} ,B^z_{i})$. Here $\mu_B$ is the Bohr magneton and $g_i$ the $g$-factor of the $i$-th dot.
The coupling rates $J_{i}^{\alpha}$ ($i=1,2$) depend on the spin-orbit coupling strength, and can be adjusted by controlling the magnetic fields or gate voltages~\cite{geyer2024anisotropic}. In the following discussions, we adopt dimensionless units for the magnetic fields $\bm{B}_i$
and coupling rates $J_{i}^{\alpha}$. Specifically, the magnetic fields and couplings are normalized by a reference energy scale (e.g. the spin-splitting energies $h_0= \mu_B g_i B_0$ with $B_0 =100$ mT) to simplify the numerical analysis and parameter tuning.


In what follows,  we shall theoretically implement high-fidelity three-qubit gates in this coupled quantum-dot system, ensuring robustness against various types of noise through the optimization of control pulses.
Unlike previous approaches~\cite{banchi2016quantum,innocenti2020supervised,lewis2024geodesic}, we utilize QAQC algorithms~\cite{khatri2019quantum} within the framework of variational quantum algorithms (VQAs)~\cite{cerezo2021variational} to produce robust shaped pulses to generate robustly shaped pulses for quantum gate construction.
A key advantage of QAQC is that the cost function is estimated directly on a quantum computer, making it well-suited for quantum system optimization. In our implementation, we use a gradient-based optimizer in a noiseless environment. In contrast, for environments with incoherent noise, a gradient-free optimizer is employed to achieve robust optimization~\cite{arrasmith2021effect}.

\subsection{VQAs}

VQAs are a promising class of hybrid quantum-classical algorithms designed to address optimization problems and simulate quantum systems. They are particularly advantageous for problems with cost functions that can be efficiently implemented as low-depth quantum circuits~\cite{broughton2020tensorflow,cerezo2021variational}.
A VQA consists of two main components: a quantum processor and a classical computer, as shown in Fig.~\ref{fig1}(a). The quantum processor prepares a quantum state, applies parameterized unitary operations, and measures the output. The measured results are processed by the classical computer, which optimizes the parameters and feeds them back to the quantum processor to minimize the cost function. This iterative process continues until a specified termination condition is met. A central component of VQAs is the ansatz, which efficiently encodes potential solutions to the problem at hand~\cite{farhi2014quantum,kandala2017hardware}. In this work, we adopt the Hamiltonian Variational Ansatz (HVA), which is particularly suited for preparing ground states of Hamiltonians composed of noncommuting terms~\cite{wiersema2020exploring}. The Hamiltonian in Eq.~\eqref{Hamiltonian} can be expressed as:
\begin{equation}
H(\bm{\theta}) = \sum^{Q}_{j=1} \theta_jH_j,
\end{equation}
where $\bm{\theta}=(\theta_1,\theta_2,\cdots,\theta_Q)$ represents the parameters of the Hamiltonian, including terms like $h^{\alpha}_i$ and $J^\alpha_i$, and $Q$ is the total number of parameters. Each $H_j$ corresponds to a single Pauli matrix or the product of two Pauli matrices. Since the terms generally do not commute ($[H_j$,$H_{j'}]\neq0$), the Hamiltonian requires careful treatment. 
In quantum computation, the time evolution of a system can be simulated using quantum gates. The Suzuki-Trotter decomposition \cite{trotter1959product,suzuki1976generalized} allows us to approximate the evolution operator $e^{-iHt}$ , even when the Hamiltonian contains noncommuting terms. The decomposition breaks the time evolution into a sequence of simpler operations. The approximate evolution operator in HVA form is:
\begin{equation}
\label{UHVA}
U_{\text{HVA}}(\bm{\theta}) = \prod_{l=1}^m \left(\prod_{j=1}^Q e^{-i\theta_j H_j t_0}\right),
\end{equation}
where $t_0=t/m$, $t$ is the total evolution time, and $m$ is the Trotter-Suzuki depth. Since  $t$ and $t_0$ are fixed, the evolution operator (\ref{UHVA}) becomes time-independent. To account for the periodicity of the exponential terms,  the parameters can be constrained to $\theta_j\in[-\pi,\pi]$.
The unitary operator is implemented using a specific ansatz circuit, as illustrated in Fig.~\ref{fig1}(c). The circuit employs single-qubit rotation gates $R_{\alpha}$ and two-qubit entangling gates $R_{\alpha,\alpha}$ defined as
\begin{eqnarray}
\label{singlegate}
R_{\alpha}(\theta_j) &=& e^{-\frac{i}{2}\theta_j \sigma_{\alpha} }, \\
\label{twogate}
R_{\alpha,\alpha}(\theta_j) &=& e^{-\frac{i}{2}\theta_j \sigma_{\alpha} \otimes \sigma_{\alpha}},
\end{eqnarray}
with $\alpha=x,y,z$. This ansatz circuit, including single-qubit rotations (X, Y, Z) and two-qubit interactions (XX, YY, ZZ) as detailed below, efficiently approximates the time evolution governed by $H$, allowing us to compare its performance with the desired target quantum gate.

\subsection{QAQC algorithm }

Designing a time-independent Hamiltonian $H(\bm{\theta})$ for high-fidelity three-qubit gates is formulated as an optimization problem, where the objective is to minimize the distance between the time-evolution operator $e^{-iH(\bm{\theta})t}$ and the target unitary operator $U_{\text{target}}$.
The cost function $C(\bm{\theta})$, which quantifies this distance, is based on the Hilbert-Schmidt inner product between the target gate $U_{\text{target}}$ and the parameterized gate sequence $U_{\text{QC}}({\bm{\theta}})$.  It is defined as~\cite{nielsen2002simple}:
\begin{equation}
C(\bm{\theta}) = 1- \frac{1}{2^{2n}}|\text{Tr}[U^\dagger_{\text{target}}U_{\text{QC}}(\bm{\theta})]|^2.
\label{cost}
\end{equation}
In this expression, the second term represents the fidelity 
$F$ of the implemented gate relative to the target gate. Consequently, the cost function $C(\bm{\theta})$ can also be interpreted as the infidelity, which we reference later in this work. By minimizing 
$C(\bm{\theta})$, we maximize the fidelity, ensuring that the implemented gate approximates the target gate with high accuracy.
The computational cost of evaluating $C(\bm{\theta})$ on a classical computer scales with the dimensions of the Hilbert space, making it resource-intensive for large systems. To address this,  the QAQC algorithm — featuring short circuit depths — is employed to compute the cost function efficiently on a quantum computer, as illustrated in Fig.~\ref{fig1}(b). The QAQC process begins with the preparation of a maximally entangled state using a combination of Hadamard and CNOT gates. The parameterized circuit, representing the HVA, $U_{QC}(\bm{\theta})=U_{\text{HVA}}(\bm{\theta})$, is applied to one half of the entangled state. While the HVA is a natural choice, alternative, potentially more efficient ansatz, e.g. counter-diabatic terms, can also be used, to improve performance and reduce the parameter number \cite{PhysRevApplied.15.024038}. To measure the overlap between the resulting state and the original maximally entangled state, the inverse circuit is applied, and the probability of observing the all-zero measurement outcome is quantified. This measurement provides an estimate of the cost function directly on the quantum processor. The optimization of 
$\bm{\theta}$ is then carried out using a classical optimizer, which iteratively updates the parameters to minimize the cost function. Depending on the noise characteristics of the environment, gradient-based optimizers can be used in noiseless settings, while gradient-free optimizers are preferred in noisy scenarios~\cite{crooks2019gradients, mitarai2018quantum, arrasmith2021effect}, see Appendix \ref{GD}.
This hybrid quantum-classical framework leverages the computational advantages of quantum hardware for cost function evaluation,  enables the design of robust, high-fidelity three-qubit gates optimized for operation in noisy quantum-dot systems. Here quantum circuits are constructed using the MindQuantum software library \cite{xu2024mindspore}, which provides tools for the development and optimization of complex quantum simulations.

\begin{figure*}[htbp]
\centerline{
\includegraphics[width=0.98\textwidth]{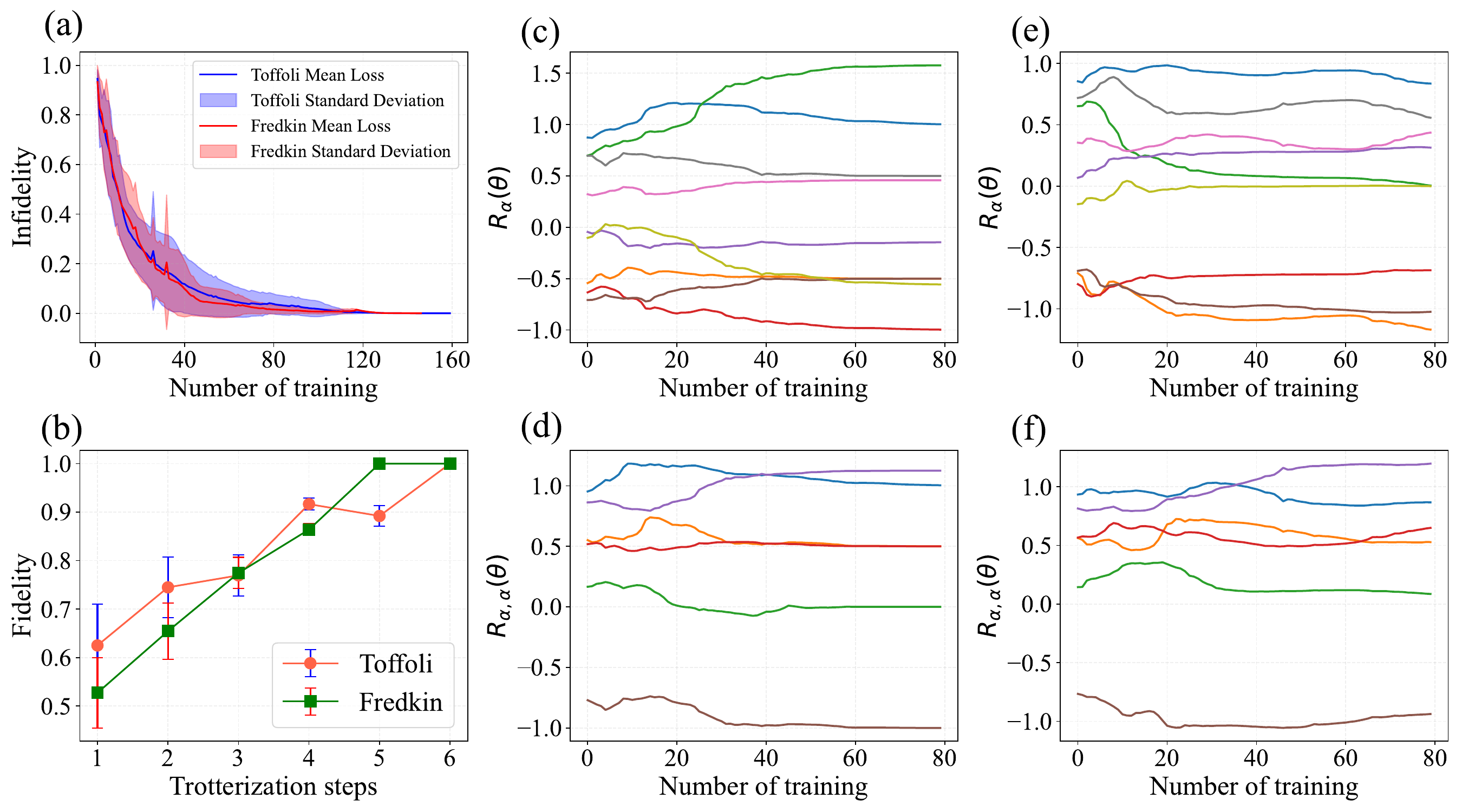}}
\caption{{(a) 
The infidelity, as defined in Eq.~\eqref{cost}, for the Toffoli and Fredkin gates is shown as a function of the number of training iterations. For each case, ten sets of randomly initialized parameters are selected, with the shaded region indicating the standard deviation of the loss during the training process.
(b) The average fidelity, along with the standard deviation, is plotted as a function of the Trotter steps $m$ for both Toffoli and Fredkin gates.
(c) and (d) depict the training dynamics of the parameters in single-qubit $R_{\alpha}(\theta_j)$ and two-qubit $R_{\alpha,\alpha}(\theta_j)$ control pulses for Toffoli gate with $m=6$. These plots summarize the optimization trajectories  for one parameter set.  
(e) and (f) illustrate the same training dynamics  as in (c) and (d), but for the Fredkin gate with $m=5$. Similar to the Toffoli gate, the initial parameters are randomly drawn from the range $[-1,1]$ with a Gaussian distribution, and the training proceeds until the fidelity threshold is achieved.}
}
\label{fig2}
\end{figure*}

\section{Preparation of the Toffoli and Fredkin gates}
\label{result}

Now, we proceed to the critical stage of optimizing the model parameters to construct robust quantum gates for the three-qubit system. Specifically, we target the implementation of the most commonly used three-qubit quantum gates: the Toffoli and Fredkin gates \cite{fredkin1982conservative}. These gates are optimized and evaluated under two distinct scenarios: a noiseless environment and a noisy environment. This dual approach allows us to rigorously test the performance and robustness of the gates against varying conditions, ensuring their practical applicability in realistic quantum-dot systems.



\subsection{Toffoli gate}

The CNOT gate operates on a two-qubit quantum register, flipping the second qubit (target qubit) if and only if the first qubit (control qubit) is in state 
$\ket{1}$. 
The Toffoli gate, also known as controlled-controlled-NOT, extends the functionality of the CNOT gate, by involving multiple control qubits. Specially, it acts as an ``AND'' gate with $N - 1$ control qubits and one target qubit. An arbitrary quantum state $\ket{\psi} = \bigotimes_{i=1}^N \ket{i}$ is transformed by 
the Toffoli gate $\hat{\tau}$ as follows:
\begin{equation}
\hat{\tau} \bigotimes_{i=1}^N \ket{i} \rightarrow \bigotimes_{i=1}^N \ket{i} \otimes \ket{\left(\prod_{k=1}^{N-1} k\right) \oplus N},
\end{equation}
where \(\oplus\) denotes addition modulo 2. For our calculation, we set $N=3$.
We use the QAQC algorithm to evaluate the cost function, see Eq.~\eqref{cost}.
In Fig.~\ref{fig2}(a), the infidelity converges after 40 iterations, as indicated by the blue solid line.
Fig.~\ref{fig2}(b) illustrates how increasing the number of Trotter steps improves the fidelity, particularly as the system transitions to higher precision with more steps. When the Toffoli gate is decomposed by using Trotter-Suzuki method with $m = 6$, the fidelity error is less than $ 10^{-4}$. Even at low Trotter steps $(m=3)$, the operator fidelity remains 80\%. 
In the noiseless environment, we employ the L-BFGS optimization algorithm to fine-tune the parameters~\cite{PhysRevResearch.6.013147}. L-BFGS is a quasi-Newton method that approximates the inverse Hessian matrix to compute the search direction, allowing for efficient handling of high-dimensional optimization problems with limited memory usage. This method is particularly well-suited for the QAQC framework, as it balances computational efficiency and convergence speed. The training results, shown in Fig.~\ref{fig2}(c) and (d), display the evolution of the single-qubit $R_{\alpha}(\theta_j)$ and two-qubit $R_{\alpha,\alpha}(\theta_j)$ control parameters, as defined in Eqs.~\eqref{singlegate} and \eqref{twogate},  during the optimization process. With 9 single-qubit and 6 two-qubit control parameters, the algorithm ensures convergence to optimal values after approximately 40 iterations. This demonstrates the efficiency and accuracy of the QAQC approach for parameter optimization in the absence of noise.

\subsection{Fredkin gate}

The Fredkin gate, also known as the controlled-SWAP gate, is a three-qubit gate that extends the concept of conditional operations in quantum computing. It consists of one control qubit and two target qubits. The gate swaps the states of the two target qubits if and only if the control qubit is \(\ket{1}\); otherwise, the target qubits remain unchanged. Mathematically, for an arbitrary quantum state \(\ket{\psi} = \ket{c}\ket{a}\ket{b}\), where \(\ket{c}\) is the control qubit and \(\ket{a}\), \(\ket{b}\) are the target qubits, the Fredkin gate \(\hat{F}\) transforms the state as:
\begin{equation} 
\hat{F} \ket{c}\ket{a}\ket{b}=\ket{c} \otimes \begin{cases} \ket{a}\ket{b} & \text{if } c = 0 \\ \ket{b}\ket{a} & \text{if } c = 1 \end{cases}.
\end{equation}
This gate is widely used in reversible computing and quantum algorithms that require conditional swapping of qubits based on the state of a control qubit. 

Fig.~\ref{fig2}(a) shows the infidelity of the Fredkin (red solid line) gates is plotted as a function of the number of training iteration.
Again, for these calculations, we initialize the parameters with 10 randomly chosen sets sampled from a Gaussian distribution. The shaded regions represent the standard deviation of the loss function over these 10 trials. In all cases, the optimization converges after approximately 40 iterations, demonstrating the algorithm's robustness. Similar to the case of the Toffoli gate, we apply L-BFSG opimizer, as it exhibits faster convergence in noiseless conditions compared to other gradient-based optimizers.  Fig.~\ref{fig2}(b), shows the average fidelity for the Fredkin gates as a function of the number of layers 
$m$, with error bars representing the standard deviation.
High fidelities are achieved for both gates with $m\geqslant4$, and the pulse parameters consistently converge to the same values across all 10 trials.
As illustrative examples, Figs.~\ref{fig2}(e) and (f)  depict the corresponding progress for the Fredkin gate, with 9 single-qubit and 6 two-qubit control parameters. These results confirm the reliability and efficiency of the proposed algorithm in achieving high fidelity and consistent optimization of quantum gates.
These results confirm the reliability and efficiency of the proposed algorithm in achieving high fidelity and consistent optimization of quantum gates.

\begin{figure}[t]
\centering
\includegraphics[scale=0.33]{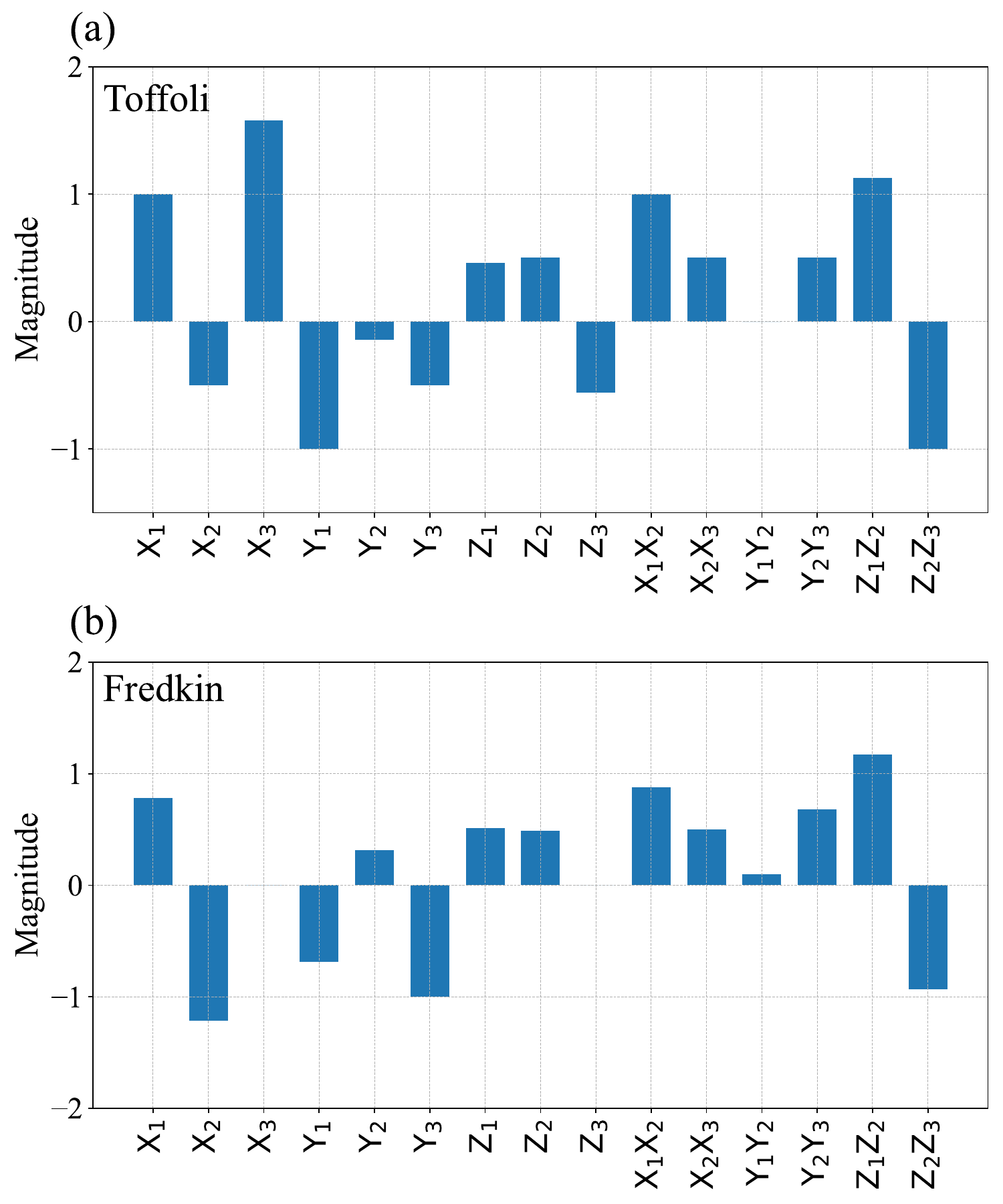}
\caption{(a) Parameter magnitudes for the Toffoli gate with $m =6$, showing the optimized values of the single-qubit and two-qubit control parameters after the optimization process converges to a fidelity threshold of $10^{-4}$. Notably, the optimization reveals that the Toffoli gate depends primarily on local field terms, with $J^y_1=0$.  
(b) Parameter magnitudes for the Fredkin gate with $m=5$, showing the optimized values of the control parameters after convergence. In contrast to the Toffoli gate, the Fredkin gate's optimization results indicate that exchange interaction terms play a more significant role, with the local field terms $h^x_3 =h^z_3=0$.}
\label{fig3}
\end{figure}



To explicitly present the optimization results, we provide the detailed values of the 15 optimized parameters for single-qubit and two-qubit gates in Fig.~\ref{fig3}. These parameters correspond to the regulated magnetic field strengths and exchange interaction terms in Eq. (\ref{Hamiltonian}), which are critical for the implementation of the Toffoli and Fredkin gates.
For clarity,  the single-qubit magnetic field strengths are denoted as  $X_1, X_2,\cdots,Z_3$, where $X_1$ respresents $h_1^{x}$, $Z_3$ corresponds to $h_3^{z}$, and so on. Moreover, the two-qubit exchange interaction terms are labeled $X_1X_2,\cdots, Z_2Z_3$, with $X_1X_2$ representing $J^x_1$, and similar conventions apply for other terms.  These detailed results in Fig.~\ref{fig3} demonstrate the different dependencies of the Toffoli and Fredkin gates. For the Toffoli gate, the optimization reveals that certain parameters, such as $J^y_1$, are not required ($J^y_1=0$), indicating a reduced dependence on specific exchange interactions. The gate primarily relies on local field strengths, as shown in Fig.~\ref{fig3}(a). In contrast, for the Fredkin gate, the results indicate $h^x_3=h^z_3=0$, highlighting the reduced significance of certain local fields for the third qubit. Instead, the exchange interactions play a more critical role in achieving high fidelity, as illustrated in Fig.~\ref{fig3}(b).
These findings provide a comprehensive view of the optimized parameters, offering guidance for the realization of three-qubit quantum gates in current quantum circuits.


\section{Discussion}
\label{discussion}

\subsection{charge noise and nuclear noise}

In the previous section, we optimize the Toffoli and Fredkin gates using QAQC algorithm in the absence of noise. However, in practical quantum systems like quantum dots, charge noise and nuclear noise pose significant challenges, affecting the coherence and control precision of qubits.

Charge noise stems from random fluctuations in charge traps and interface states. Particularly, charge traps can capture and release charge carriers, leading to random variations in the electric field near the quantum dot~\cite{zhang2015dynamic}.These variations induce instabilities in the quantum dot's potential, leading to fluctuations in the tunneling coupling $J_i^{\alpha}$, i.e.,
$J_i^{\alpha} \rightarrow J_i^{\alpha} + \delta J_i^{\alpha}$, which directly affects the interaction strength between the qubits~\cite{kuhlmann2013charge,PhysRevLett.110.140502}. The source of these charge fluctuations lies in imperfections at the semiconductor-insulator interface, where defect states create random potential fluctuations. These charge noises are particularly pronounced in materials where quantum dots are embedded in semiconductor substrates, and they can lead to significant errors in quantum operations if not properly mitigated.

Nuclear noise, on the other hand, arises due to the thermal fluctuations and interactions of nuclear spins surrounding the quantum dots. In semiconductor materials, electron spins are subject to hyperfine interactions with the local nuclear spins, which creates local magnetic fields that vary over time~\cite{chekhovich2013nuclear}. These fluctuating magnetic fields result in random phase shifts and dephasing of the electron spin states~\cite{PhysRevLett.109.140403}, leading to decoherence and a reduction in the qubits' coherence time. This type of noise is especially significant in materials with a high concentration of nuclear spins, as the dynamic changes in these spins induce noise in the electron spin qubits. For simplicity, we assume that the hyperfine interaction with nuclear spins primarily affects the Zeeman energies of separated quantum dots. This influence can cause variations in the magnetic field experienced by each qubit, effectively modifying the local magnetic field strength $h_i^{z}$. Consequently, the hyperfine coupling introduces a perturbation of the order   $h_i^{z} \rightarrow h_i^{z} + \delta h_i^{z}$,
where $\delta$ represents the fluctuation induced by the nuclear spin environment.

Furthermore, we emphasize that noise is incorporated into the system's evolution only after the compilation process. For simplicity, we assume that all three qubits are subjected to the same noise profile, ensuring a uniform treatment of noise across the system. To preserve the intrinsic properties of the exchange interactions, we impose the condition that the noise contributions remain non-negative. This assumption ensures consistency with the physical nature of the coupling strengths, allowing for realistic modeling of noise effects in the quantum dot system.

\begin{figure}[t]
\centering
\includegraphics[scale=0.33]{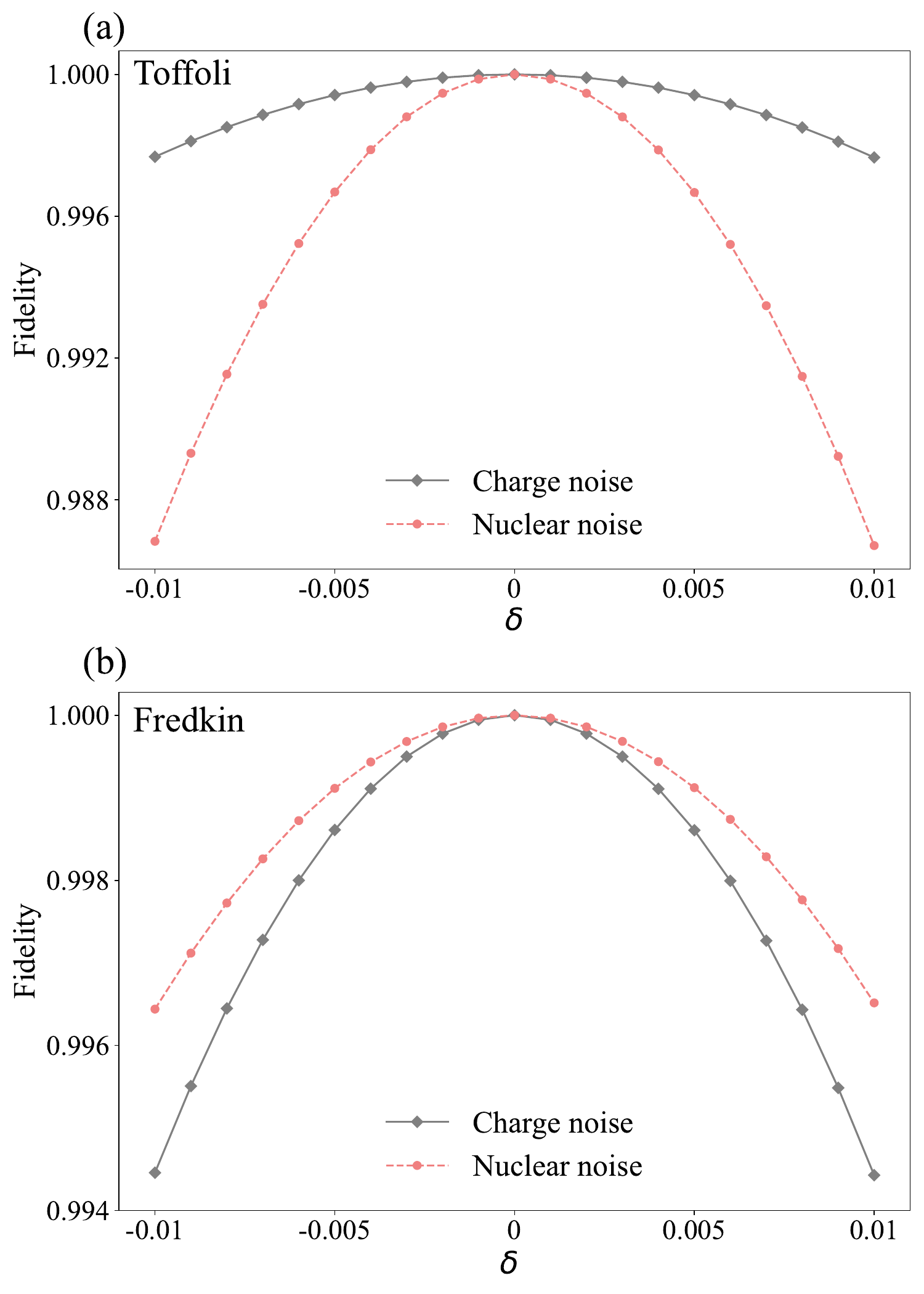}
\caption{The fidelity of (a) the Toffoli gate and (b) the Fredkin gate under the influence of charge and nuclear noise with amplitude $\delta$. The gray solid lines correspond to the fidelity affected by charge noise, while the red dotted lines show the fidelity under nuclear noise. }
\label{fig4}
\end{figure}

As shown in Figs.~\ref{fig4}(a) and \ref{fig4}(b), the parameters of the Toffoli gate and Fredkin gate optimized by the QAQC algorithm exhibit strong robustness against noise within a reasonable range. 
This robustness stems from the intrinsic capabilities of the VQAs to recover from noise and maintain high performance. Generally, variational algorithms, such as QAQC, are highly adaptable due to their ability to minimize a cost function through iterative optimization. One key aspect of their robustness lies in their capacity to compensate for coherent errors, such as over- or under-rotation of qubit states. These errors, typically caused by small inaccuracies in pulse shaping or interaction strengths, are corrected during the optimization process as the algorithm adjusts parameters to minimize the cost function. VQAs are inherently deformable, meaning they can adapt to changes in the system dynamics without compromising the ability to find optimal solutions. Importantly, certain types of noise, particularly those that only shift the location of the minimum of the cost function, do not prevent the algorithm from achieving convergence. For example, a coherent error might cause the optimization landscape to be translated or rotated slightly in parameter space, but the variational method can still locate the new minimum effectively. This capability ensures that the optimized parameters remain reliable even when the system is subject to mild disturbances or perturbations~\cite{mcclean2016theory,PhysRevX.6.031007,PhysRevX.8.011021}. Furthermore, the observed differences in the noise impacts on the Toffoli and Fredkin gates provide additional insights. The Toffoli gate demonstrates greater sensitivity to nuclear noise, as its parameters depend more heavily on local magnetic fields. Fluctuations in these fields, often induced by hyperfine interactions with surrounding nuclear spins, result in decoherence and random phase shifts that affect the optimization process. On the other hand, the Fredkin gate shows a stronger sensitivity to charge noise due to its greater reliance on exchange interactions. Charge noise, which originates from fluctuations in the electric field near quantum dots, can perturb these exchange interactions, leading to variations in the gate's performance. These findings highlight the practical advantages of QAQC algorithm in mitigating noise-related challenges. Their adaptability and error-compensation capabilities make them well-suited for applications in the NISQ era, where quantum devices are inherently noisy and require robust control schemes to ensure reliable operation.

\begin{figure}[t]
	\centering
	\includegraphics[scale=0.37]{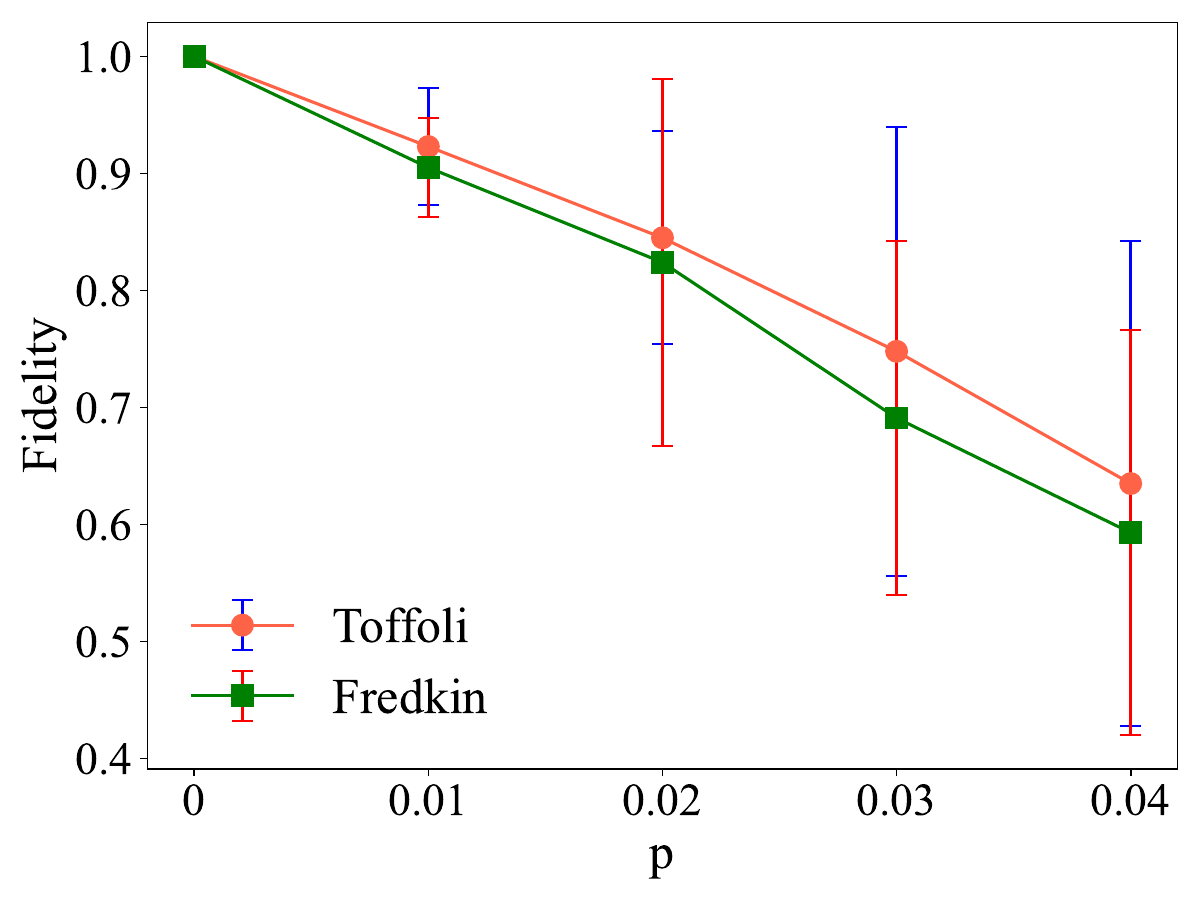}
	\caption{The average fidelity as a function of the noise parameter $p$ in Eq.~(\ref{noise}) under amplitude-damped noisy channels. The error bars indicate the standard deviation, reflecting the variability across 100 independent training sessions. Optimization is performed using the Nelder-Mead optimizer.}
	\label{fig5}
\end{figure}

\subsection{quantum device noise}

In reality, quantum computers cannot be entirely isolated from their environment, and interactions with the surrounding environment often result in decoherence of the qubits. This makes it essential to incorporate noise models that account for non-ideal unitary evolution. In this section, we analyze the performance of the QAQC algorithm when quantum states are subject to noisy channels, focusing on amplitude-damping channels, $\mathcal{N}_p^{\text{amp}}(\rho)$. These channels describe energy dissipation processes where qubits transition from an excited state to the ground state. The noise channel is defined as~\cite{PhysRevA.54.2614}:
\begin{equation}
    \mathcal{N}_p^{\text{amp}}(\rho) = E_0 \rho E_0^{\dagger} + E_1 \rho E_1^{\dagger},
\end{equation}
where the Kraus operators are given by
\begin{equation}
\label{noise}
    E_0 = \begin{bmatrix}
        1 & 0 \\
        0 & \sqrt{1-p} \\ 
    \end{bmatrix},E_1 = \begin{bmatrix}
        0 & \sqrt{p} \\
        0 & 0 \\ 
        \end{bmatrix}.
\end{equation}
Here, $p$ is the probability of decay, which characterizes the transition of the qubit from the excited state to the ground state.

In quantum circuits influenced by incoherent noise, direct gradient computation becomes challenging, necessitating the use of gradient-free optimization methods. In our analysis, we employed the Nelder-Mead optimizer, conducting 100 fidelity evaluations at each noise level $p$. The impact of amplitude-damping channels on the fidelity of quantum gates is shown in Fig.~\ref{fig5}, where red and green dots represent the performance of the Toffoli and Fredkin gates, respectively.
At a noise level of $p=0.01$, the average fidelity of the Toffoli gate is approximately $92\%$, while the Fredkin gate achieves an average fidelity of $90\%$. However, as the noise level increases to $p=0.02$, the fidelities of both gates decline to around $80\%$. Despite the reduction in fidelity with increasing noise, these results demonstrate that the QAQC algorithm exhibits good resilience to noise.

\subsection{barren plateaus}

Next, we address the barren plateaus problem~\cite{mcclean2018barren}, a significant challenge in training parametric quantum circuits. This issue arises when the gradient of the loss function (or objective function) becomes negligibly small in certain regions of the parameter space, leading to extremely slow or even ineffective parameter updates during optimization. This phenomenon, often referred to as the flat-zone problem, can cause variational algorithms to stagnate, making it difficult to converge to an optimal solution. In the barren plateaus regime, the optimization landscape flattens, and gradients become exponentially small as the number of qubits increases. This scaling issue is particularly problematic for high-dimensional quantum systems, as it limits the practical utility of variational quantum algorithms for such systems. Barren plateaus are influenced by circuit design, initialization, and the nature of the problem being solved, making their mitigation a crucial focus in quantum algorithm development.

However, our QAQC algorithm does not suffer from  the barren plateaus problem. The quantum gate 
$U_{\text{QC}}$ in our implementation, which acts on 
$n$ qubits, is an element of the unitary group $U(N)$, where 
$N=2^n$. By disregarding the global phase, $U_{\text{QC}}$ 
belongs to the special unitary group $SU(N)$, which exhibits the symmetry. This symmetry plays a crucial role in maintaining a structured optimization landscape, reducing the likelihood of encountering barren plateaus. Moreover, the specific structure of the quantum gates and the problem formulation ensures that the optimization remains effective. The inherent symmetries in the parameterized ansatz and the variational algorithm contribute to a well-behaved loss function, preventing the optimization process from being trapped in flat zones \cite{kazi2024analyzing}. Thus, the training process for our quantum circuits is robust, avoiding the stagnation associated with barren plateaus, and facilitating the efficient discovery of optimal solutions.

To illustrate this, we consider the implementation of the Toffoli and Fredkin gates using  Trotter steps $m=6$ and $m=5$, respectively. In these configurations, the algorithm optimizes 15 parameters associated with single-qubit and two-qubit gate controls. The relatively shallow circuit depth reduces the expressibility of the ansatz, thereby limiting the number of parameters. This ensures that the optimization landscape remains well-structured and avoids the barren plateaus problem~\cite{PRXQuantum.3.010313}. As the number of parameters increases, one might expect the risk of barren plateaus to grow due to the potential over-expressibility of the ansatz. However, the structured design of the QAQC algorithm mitigates this risk by preserving problem-specific symmetries and tailoring the ansatz to the system under consideration. These features ensure that expressibility is controlled and trainability is maintained, even for more complex systems. This balance between parameter complexity and algorithmic design ensures scalability while maintaining optimization efficiency.

\section{conclusion}
\label{conclusion}


To summarize, we have utilized the QAQC algorithm  to identify time-independent Hamiltonians capable of implementing three-qubit gates in quantum-dot systems. This algorithm significantly reduces the computational cost of evaluating the loss function on a classical computer while minimizing the risk of being trapped in local optima. Our approach, characterized by short circuit depths, demonstrates robust performance against charge and nuclear noise in quantum dot systems, achieving a high fidelity of approximately $99\%$. Additionally, we analyze the impact of quantum device noise, such as amplitude-damping channels, showing that the gates maintain high performance within a reasonable noise threshold. This robustness highlights the algorithm's potential for NISQ devices.

A key different of our method lies in its reliance on time-independent Hamiltonians, unlike other hybrid classical-quantum algorithms for quantum control \cite{PhysRevLett.125.260511,PhysRevResearch.3.023165,PhysRevResearch.5.023173,sun2022optimizing}, which typically optimize time-dependent Hamiltonians. Time-independent Hamiltonians offer a simpler implementation in experimental setups and reduce the complexity of pulse engineering, while still achieving comparable or superior performance.

Finally, our scheme can be extended to other physical platforms, such as superconducting qubits \cite{reed2012realization,fedorov2012implementation}, trapped ions \cite{PhysRevLett.102.040501}, and silicon photonic chips \cite{li2022quantum}. However, considering the current limitations in the number of qubits compared to superconducting qubits and trapped-ion systems, the design and optimization of high-fidelity three-qubit gates remain a crucial step in advancing the capabilities of quantum dot-based quantum computing. Furthermore, our approach can be complemented by machine learning \cite{banchi2016quantum,PhysRevApplied.6.054005,PhysRevApplied.22.024009} and digitized counter-diabatic ansatz  (including XY and YZ terms) \cite{PhysRevApplied.15.024038} to reduce the quantum gate number and circuit depth further. Yet, our results have demonstrated the flexibility and power of the QAQC algorithm for constructing Hamiltonians, providing a versatile pathway for realizing multi-qubit quantum gates and contributing to the development of scalable, high-fidelity quantum computing platforms across various architectures.

\section*{Acknowledgment}
Y.Z and F.P acknowledge OIST for hospitality.
This work was supported by the National Natural
Science Foundation of China (Grants Nos. 12075145, 112374247 and 11974235), the Shanghai Municipal Science and Technology Major Project (Grant No.2019SHZDZX01-ZX04), and the Innovation Program for Quantum Science and Technology (Grant No. 2021ZD0302302).  This work was also supported by the Japan-China Scientific Cooperation Program between JSPS and NSFC Under Nos. 120227414 and 12211540002. X.C. also appreciates the support from  the project grant PID2021-126273NB-I00 funded by MCIN/AEI/10.13039/501100011033 and by ``ERDF A way of making Europe" and ``ERDF Invest in your Future", and the Spanish Ministry of Economic Affairs and Digital Transformation through the QUANTUM ENIA project call-Quantum Spain project. 

\appendix

\section{Classical optimizers}
\label{GD}

\subsection{Gradient-based optimization}
Gradient-based optimization methods leverage the gradient of the loss function with respect to the model parameters to iteratively adjust those parameters in the direction that minimizes the loss, thereby improving the model's performance.

One of the most commonly used gradient-based optimizers is gradient descent, which updates the parameters as follows:
\begin{equation} \bm{\theta}_{k+1} = \bm{\theta}_k - \alpha \cdot \nabla_{\theta} J(\bm{\theta}), \end{equation}
where  $\alpha$ is the learning rate, and $J(\bm{\theta})$ is the function to be optimized.
In this work, we utilize the L-BFGS (Limited-memory Broyden-Fletcher-Goldfarb-Shanno) optimizer~\cite{liu1989limited}, a variant of the BFGS optimizer~\cite{head1985broyden}. These algorithms belong to the class of quasi-Newton methods, which aim to find local minima or maxima of a function by iteratively approximating the inverse Hessian matrix.
The L-BFGS optimizer offers several advantages. For instance, it uses gradient information to approximate the Hessian matrix iteratively. Unlike the original BFGS method, L-BFGS does not store the full Hessian matrix, reducing memory requirements significantly. Its limited-memory approach makes it well-suited for optimizing high-dimensional functions where explicitly computing or storing the full Hessian is computationally prohibitive. Therefore,
by efficiently balancing gradient information and memory usage, L-BFGS enables the optimization of complex loss landscapes with large numbers of parameters.


\subsection{Gradient-free optimization}
\label{GF}

Gradient-free optimization methods do not require the computation of gradients for the objective function, making them especially useful for scenarios where the function is non-differentiable, discontinuous, noisy, or computationally expensive to evaluate. Common examples of gradient-free optimizers include the Nelder-Mead method, Constrained Optimization by Linear Approximations (COBYLA), and evolutionary algorithms.

In this work, we employ the Nelder-Mead optimizer, a derivative-free, iterative method that minimizes a function by constructing a simplex—a geometric structure $n+1$ points for $n$-dimensional parameter space. The algorithm iteratively adjusts the simplex to converge on the function's minimum. Its derivative-free nature makes it particularly suitable for handling noisy or non-smooth functions, as encountered in quantum gate optimization under realistic noise models.

By using both gradient-based and gradient-free optimization techniques, our approach achieves flexibility and robustness, addressing a wide range of challenges in quantum gate optimization. This combination ensures efficient convergence, even in highly complex or noisy optimization landscapes. For a comprehensive discussion of classical optimizers in VQAs, refer to Ref.~\cite{PhysRevResearch.6.013147}.


\bibliography{ref} 

\end{document}